\documentclass{book}    % fundamental class file is 'book'.
\usepackage{piers}  % use the file 'piers.sty'.
\usepackage{graphicx}
\begin{document}

%=== Input Title here ================================
\title{Comparative study of multipactor effect in rectangular and \\parallel-plate waveguides partially loaded with dielectric}
\maketitle
%===================================================

%=== List of authors (in order) ========
%-- Author(s) for the first affiliation ---
\author      {A. Berenguer}
\affiliation {Departamento de Ingenier\'ia de Comunicaciones, Universidad Miguel Hern\'andez de Elche}
\address     {Avenida de la Universidad s/n}% optional
\city        {Elche}
\postalcode  {03202}% optional
\country     {Spain}
\phone       {}    % optional
\fax         {}    % optional
\email       {aberenguer@umh.es}  % optional
\misc        { }  % optional
\nomakeauthor
%------------------------------------

%=== List of authors (in order) ========
%-- Author(s) for the second affiliation ---
\author      {A. Coves}
\affiliation {Departamento de Ingenier\'ia de Comunicaciones, Universidad Miguel Hern\'andez de Elche}
\address     {Avenida de la Universidad s/n}% optional
\city        {Elche}
\postalcode  {03202}% optional
\country     {Spain}
\phone       {+34966658415}    % optional
\fax         {}    % optional
\email       {angela.coves@umh.es}  % optional
\misc        { }  % optional
\nomakeauthor
%-------------------------------------
%=== List of other authors (in order) ========
\author      {E. Bronchalo}
\affiliation {Departamento de Ingenier\'ia de Comunicaciones, Universidad Miguel Hern\'andez de Elche}
\address     {Avenida de la Universidad s/n}% optional
\city        {Elche}
\postalcode  {03202}% optional
\country     {Spain}
\phone       {+34966658593}    % optional
\fax         {}    % optional
\email       {ebronchalo@umh.es}  % optional
\misc        { }  % optional
\nomakeauthor

%---Output of Authors----------------------
\begin{authors}

{\bf A. Berenguer}$^{1}$, {\bf A. Coves}$^{1}$, {\bf and E. Bronchalo}$^{1}$\\
\medskip
$^{1}$Departamento de Ingenier\'ia de Comunicaciones. Universidad Miguel Hern\'andez de Elche, Spain
\end{authors}
%--------------------------

% correct bad hyphenation here
\hyphenation{wave-guide simu-lations a-ppearing}

%---Content of Paper Abstract-----------------------
\begin{paper}

\begin{piersabstract}
In this work, a comparative study of the susceptibility chart in a partially dielectric-loaded rectangular waveguide and in its equivalent partially dielectric-loaded parallel-plate waveguide with the same height is performed. This study shows how the inhomogeneity of the electric field inside the rectangular waveguide modifies the edges of the multipactor susceptibility chart with respect to that predicted by the parallel-plate waveguide case. Moreover, comparisons of the evolution of the population and the trajectory of the effective electron and the DC field appearing in the waveguide in a point well inside the multipactor region obtained with both models are performed, showing some quantitative differences in the multipactor evolution, but a similar qualitative behaviour is observed.
\end{piersabstract}

%---Content of Paper Text-----------------------
\psection{Introduction}
The multipactor effect is an electron discharge that may appear in particle accelerators and microwave devices such as waveguides in satellite on-board equipment under vacuum conditions. Multipactor research lines are aimed to study and characterize the phenomenon to predict the conditions for its appearance\cite{ar:Esteban}, \cite{ar:Quesada}, taking advantage of available susceptibility charts in empty parallel-plate waveguides obtained with analytical models \cite{ar:Hatch}, and they are directly used to predict multipactor breakdown in the component under study, which is going to happen in the point of highest field intensity. However, many RF devices, such as filters, multiplexers, and RF satellite payloads, include dielectric materials commonly employed as resonators and supporting elements. In previous works \cite{ar:Torregrosa06}–-\cite{ar:Torregrosa10}, a theoretical model for studying the multipactor effect in a parallel-plate dielectric-loaded waveguide has been developed, which can be used to obtain susceptibility charts in partially dielectric-loaded waveguide components. Nevertheless, such susceptibility diagrams do not take into account the non-uniform nature of the electromagnetic fields inside the waveguide. For this reason, the authors have recently developed a theoretical model for the analysis of the multipactor effect in a partially dielectric-loaded rectangular waveguide \cite{ar:berenguer19}. This model has been used in this work to compute the susceptibility chart in a partially dielectric-loaded rectangular waveguide, and these results have been compared with that of its equivalent partially dielectric-loaded parallel-plate waveguide with the same height, obtained with a parallel-plate 3D multipactor model described in \cite{ar:berenguer15}. The obtained results show that the inhomogeneity of the electric field inside the rectangular waveguide modifies the limits of the multipactor susceptibility chart with respect to that predicted by the parallel-plate waveguide case. However, a comparison of the evolution of the population and the trajectory of the effective electron and the DC field appearing in the waveguide with both models, in a point of the susceptibility chart well inside the multipactor region, shows some quantitative differences in the multipactor evolution, although they finally yield a similar qualitative behaviour.

\psection{Comparison of the Multipactor Susceptibility Chart in a Rectangular and its Equivalent Parallel-Plate Waveguide Partially Loaded with a Dielectric layer}
\begin{figure}[t]
	\centerline{\includegraphics[width=0.75\textwidth]{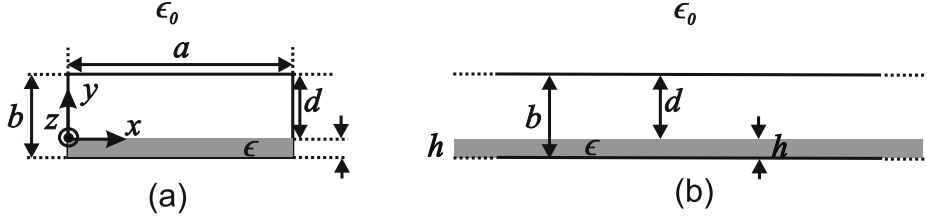}}
	\caption{Cross section of a dielectric-loaded rectangular waveguide (a), and of its equivalent parallel-plate dielectric-loaded waveguide (b).}
	\label{fig:esquema}
\end{figure}
In this section, a recently developed simulation model \cite{ar:berenguer19}, \cite{ar:berenguer16} for analysing the multipactor effect appearing in a rectangular waveguide partially loaded with a dielectric layer, when propagating high power resonant RF fields and under vacuum conditions, has been used to compute the susceptibility chart in the waveguide. Unlike the problem of studying the multipactor effect in a partially dielectric-loaded parallel-plate waveguide, in which the effective electron trajectories can be described with analytical formulas \cite{ar:Coves08},\cite{ar:berenguer15}, the greater complexity of numerically solving the differential equations of the effective electron trajectories in the inhomogeneously filled rectangular waveguide makes it necessary to adopt a series of simplifications in our model. In particular, a single effective electron model has been adopted, and, in addition, space charge effects have not been considered. Nevertheless, on the one hand, the RF fields inside the waveguide have been considered without approximations. And, on the other hand, the growth of an electrostatic field $\textbf{E}_\text{DC}$ in the waveguide during the multipactor process due to the appearance of charges located at different positions on the dielectric surface, associated to the absorption or emission of electrons in the impact locations, has also been considered to accurately obtain the trajectory of the electrons inside the waveguide, considering both the spread in secondary emission energy and the angle of the secondary electrons after each impact on the waveguide walls. This assumption has proved to  properly account for the charging of the dielectric material, given that the discharging time for dielectrics is much higher than the typical time for a multipactor discharge. A detailed explanation of the model has been omitted here for the sake of conciseness, and it can be found in \cite{ar:berenguer19}, \cite{ar:berenguer16}.

Fig.\ref{fig:esquema}(a) shows the cross section of a partially dielectric-loaded rectangular waveguide of width~$a$ and height~$b$, loaded with a dielectric slab of relative permittivity~$\epsilon_r$ and dimensions $a\times h$ placed over the bottom waveguide wall, being $d$ the empty waveguide height where the effective electron travels. We have analyzed the multipactor effect in a nonstandard silver-plated rectangular waveguide of width $a = 19.05$\,mm and height $b = 0.4$\,mm, in which a thin dielectric layer of thickness $h=0.025$\,mm and $\epsilon_r = 2.1$ (corresponding to a realistic dielectric film commonly used in space  applications) has been placed over the bottom surface of the waveguide, being $d =  b-h$ the vertical air gap in the waveguide. We have restricted our study to the monomode regime in the partially dielectric-loaded rectangular waveguide, where only the fundamental mode, TM$^{y}_{10}$, propagates \cite{Harrington}. A study of the susceptibility chart of this waveguide has been performed and compared with results obtained in its equivalent parallel-plate waveguide partially loaded with the same dielectric layer (see Fig. \ref{fig:esquema}(b)), where the charge appearing on the dielectric surface is considered to be uniformly distributed on an area $A$. We want to highlight the effect of the inhomogeneity of both the RF fields and charge distribution appearing on the dielectric surface in the rectangular waveguide with respect to the parallel-plate one. For the comparison, the multipactor simulation model developed in \cite{ar:berenguer15} for a parallel-plate waveguide including the 3D motion of the electrons has been employed. Although this model is able to consider both multiple effective electrons and space charge effects, they have been neglected here for the comparison. In the vertical axis of the susceptibility chart, it is plotted an effective voltage, $V_\text{eff}$, which in the rectangular waveguide has been numerically calculated as the line integral of the $E_{y}$ component of the electric field (evaluated at the center of the waveguide $x = a/2$) from $y_{1} = 0$ to $y_{2} = d$. In the equivalent parallel-plate waveguide, this voltage is obtained as $E_0\times(d-h)$, being $E_0$ the RF field amplitude, which in this case can be calculated as
\begin{equation}
E_{0}=\frac{V_{0}\epsilon_{r}}{h+\epsilon_{r}(d-h)}
\end{equation}
In the horizontal axis, it has been plotted the product $f \times d$. This susceptibility chart is only applicable to these waveguides, and not to any waveguide with an air gap $d$, given that the electromagnetic field distribution depends on the characteristics of the dielectric layer (i.e., on its dimensions and relative permittivity). To obtain the susceptibility chart, the simulation is run for 100~RF cycles at each $V_\text{eff}$ and $f \times d$ point, and repeated a sufficient number of times (typically, a higher number is needed at the multipactor zone edges than inside or outside of them). In each simulation, the effective electron is launched at $x=a/2$, $z=0$, at a random position $y_{0}$ in the $y$ axis between $y=0$ and $y=d$, following a cosine distribution of the polar angle, and with a departure kinetic energy following a probability density function given in \cite{ar:Scholtz}. In the parallel-plate case, the charge appearing on the dielectric surface has been considered to be uniformly distributed on an area $A=10$ cm$^{2}$. Using all simulations, the multipactor discharge is assumed to have occurred when the arithmetic mean of the final population of electrons is greater than~1. Additionally, given that the  emission or absorption of electrons by the dielectric surface gives rise to an increasing DC field in the waveguide that eventually may turn off the discharge \cite{ar:berenguer19}, a minimum mean value of the remaining $E_\text{DC}$ field in the waveguide using all simulations is also used as an additional criterion to assume that a multipactor discharge has occurred at a given $V_\text{eff}$ and $f \times d$ point.

In the susceptibility chart shown in Fig. \ref{fig:carta}, the multipactor discharge zones in the partially dielectric-loaded rectangular waveguide have been represented with black points, while the gray points correspond the multipactor discharge region in the partially dielectric-loaded parallel-plate waveguide. At the sight of this figure, it can be checked that in the observed multipactor regions in the analyzed $V_\text{eff}$ and $f \times d$ ranges, a higher multipactor threshold is predicted in the partially dielectric-loaded rectangular waveguide. Moreover, this difference is constant for all values of $f \times d$. And, on the contrary, the upper limit of the multipactor region is lower for the partially dielectric-loaded rectangular waveguide than for the partially dielectric-loaded parallel-plate waveguide. In order to better understand these results, in the next section, the multipactor evolution at different regions of the susceptibility chart obtained with both models is analyzed. 
\begin{figure}[htbp]
	\centerline{\includegraphics[width=0.60\textwidth]{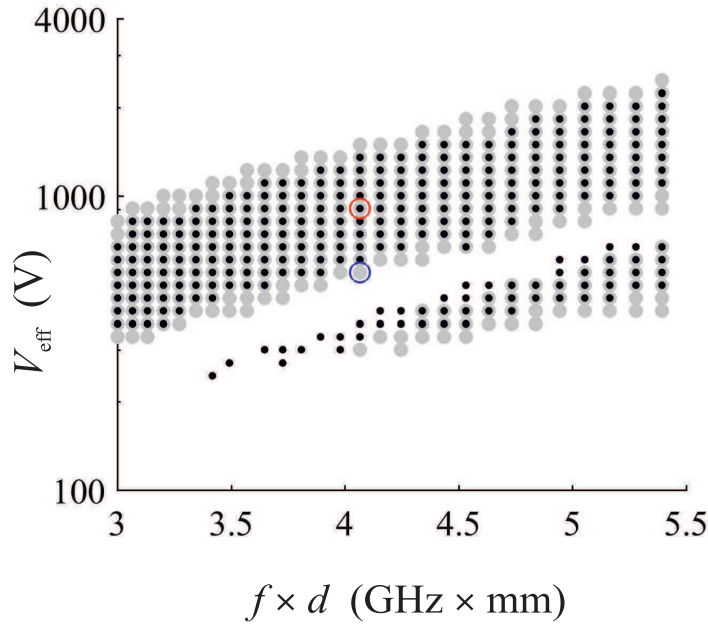}}
	\caption{Susceptibility chart of a partially dielectric-loaded rectangular waveguide (black points) and of its equivalent parallel-plate waveguide (gray points). The point $f\times d=4.01$\,GHz$\cdot$mm and $V_\text{eff}= 908$\,V (well inside the first order multipactor region in both waveguides) is highlighted in red, while the point $f\times d=4.01$\,GHz$\cdot$mm and $V_\text{eff}= 550$\,V (in the lower limit of the first order multipactor region) is highlighted in blue.}
	\label{fig:carta}
\end{figure}

\psection{Multipactor Evolution at Different Regions of the Susceptibility Chart}
With the purpose of having a better understanding of the differences of the multipactor evolution in the equivalent partially dielectric-loaded rectangular and parallel-plate waveguides, several simulations of the multipactor evolution at different regions of the susceptibility chart obtained with both models are performed and analyzed next. Simulations assuming different initial phases of the RF field have been done, and similar results were obtained. 

A first point well inside the first order multipactor region in both waveguides has been chosen, corresponding to $f\times d=4.01$\,GHz$\cdot$mm and $V_\text{eff}= 908$\,V (highlighted in red in Fig. \ref{fig:carta}). In Figs. \ref{f:rect908} and \ref{f:gppp908} the evolution of the following magnitudes in each case have been represented: total number of electrons $N$ (black line), $E_{y,DC}$ at the electron position (blue line), $E_{y,\text{RF}}$ (red line), and electron position $y$ (gray line). The other DC and RF fields components have turned out to be several orders of magnitude lower than the represented ones in the cases under study and have not been displayed.
\begin{figure}[htbp]
	\centerline{\includegraphics[width=0.8\textwidth]{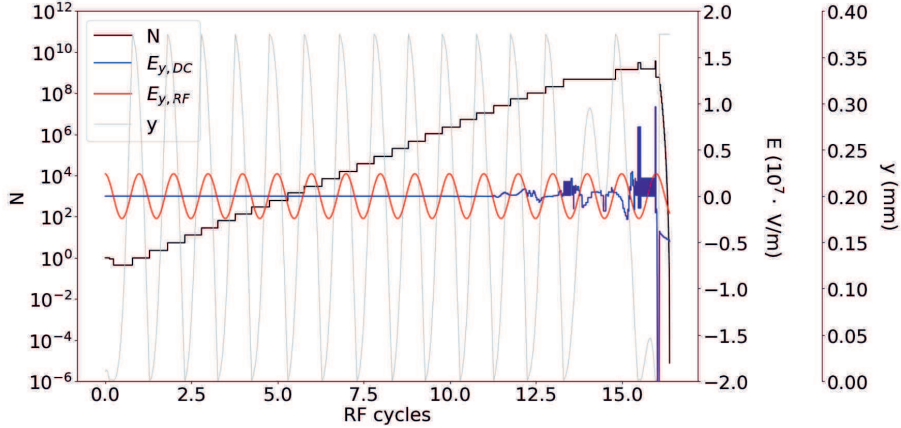}}
	\caption{Multipactor evolution in the partially dielectric-loaded rectangular waveguide at $f\times d=4.01$\,GHz$\cdot$mm and $V_\text{eff}= 908$\,V: total number of electrons $N$ (black line), $E_{y,DC}$ at the electron position (blue line), $E_{y,\text{RF}}$ (red line) and electron position $y$ (gray
line).}
	\label{f:rect908}
\end{figure}	
In Fig. \ref{f:rect908}, it is plotted the multipactor evolution in the partially dielectric-loaded rectangular waveguide at the point $f\times d=4.01$\,GHz$\cdot$mm and $V_\text{eff}= 908$\,V as a function of the time normalized to the RF period. Given that resonance conditions are met at this point (in the center of the 1st multipactor region), once the effective electron is synchronized with the RF field after a few RF cycles, the $y$ coordinate of its trajectory (represented with gray line) shows collisions with the top metallic and bottom dielectric surface in a first-order multipactor process until the 13-th RF cycle, remaining in the vicinity of $x=a/2$ and $z=0$ ---given that the electron has nearly no acceleration in such directions. In this first stage of the multipactor evolution, the total number of electrons $N$ (black solid line) follows an exponential growth. This progressive growth of $N$ entails the appearance of charges on the dielectric surface, whose value is proportional to the emitted or absorbed electrons in each impact on this surface. Such charges give rise to the appearance of an electrostatic field in the empty gap. Once the population of electrons reaches a significant number ($N\approx 10^{8}$ in the conditions under study), the $y$-component of the DC field, $E_{y,\text{DC}}$ (which has been plotted in~Fig. \ref{f:rect908} at the electron position with blue line) becomes comparable to $E_{y,\text{RF}}$ (red line), and the effective electron loses its previous multipactor synchronization. From this moment on, the electrons collide with the top metallic or bottom dielectric surface much sooner or later than the instants when the RF electric field changes its sign, which implies low impact energy collisions, and consequently, absorption of electrons, yielding to the appearance of greater charges on the dielectric surface, and contributing to a higher DC field acting on the waveguide. A high negative value of this field near the dielectric surface finally results in a single-surface multipactor regime in the metallic surface (see the $y$ position of the electron from RF cycle $16$), with successive low impact energy collisions, until the discharge turns off. The remaining DC field in the waveguide at the end of the simulation is the proof that a discharge has occurred.
\begin{figure}[htbp]
	\centerline{\includegraphics[width=0.8\textwidth]{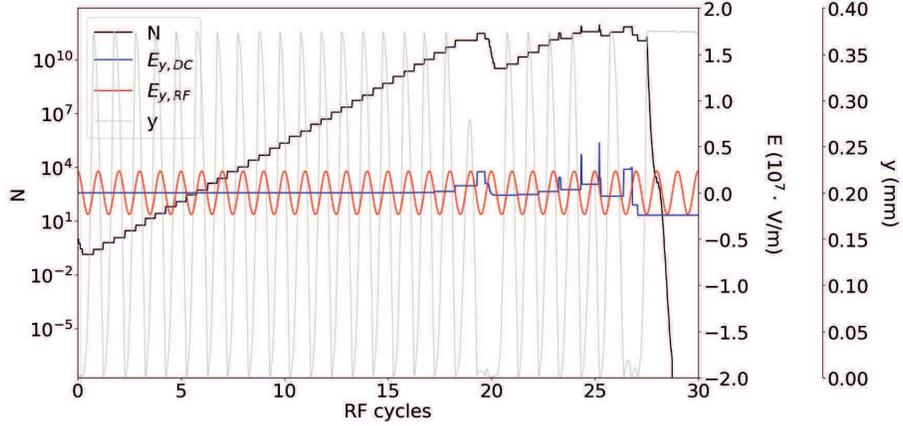}}
	\caption{Multipactor evolution in the partially dielectric-loaded parallel-plate waveguide at $f\times d=4.01$\,GHz$\cdot$mm and $V_\text{eff}= 908$\,V: total number of electrons $N$ (black line), $E_{y,DC}$ at the electron position (blue line), $E_{y,\text{RF}}$ (red line) and electron position $y$ (gray
line).}\label{f:gppp908}
\end{figure}

Equivalent results are plotted in Fig. \ref{f:gppp908} for the partially dielectric-loaded parallel-plate waveguide at the same point $f\times d=4.01$\,GHz$\cdot$mm and $V_\text{eff}= 908$\,V, which also corresponds to resonance conditions of the effective electron with the RF field. This can be checked through the exponential growth followed by the total number of electrons $N$ (black solid line) until RF cycle $17$. In the same way, the dielectric surface of the waveguide charges proportionally to the emitted or absorbed electrons by it in each impact, giving rise to an increasing electrostatic field -which in this case is assumed to be constant in the empty gap-. Once $E_{y,\text{DC}}$ (blue line) becomes comparable to $E_{y,\text{RF}}$ (red line), the synchronization of the effective electron with the RF field is lost, and after a few RF cycles, the appearance of a negative DC field pushes the electrons to the upper metallic surface, and successive low impact energy collisions take place in a single-surface multipactor regime, anew turning off the discharge. Again, the DC field remaining in the waveguide proofs that there has been a multipactor discharge. Therefore, a qualitatively similar behaviour is observed with both simulation models at this point, and only quantitative differences in the maximum population of electrons $N$ reached and in the final DC field are found, which are basically due to the inhomogeneity of the charge distribution appearing on the dielectric surface in the rectangular waveguide (instead of being uniformly distributed in a given area). 
 
A different point of the susceptibility chart in the lower limit of the first order multipactor region is next analyzed, corresponding to $f\times d=4.01$\,GHz$\cdot$mm and $V_\text{eff}= 550$\,V. Indeed, no multipactor discharge has been observed in the rectangular waveguide at this point (although 360 simulations have been performed at this point, none of them has shown a growth of the population of electrons). This fact can be understood at the sight of one of such simulations of the multipactor evolution in the rectangular waveguide, represented in Fig. \ref{f:rect550}, where although the $y$ coordinate of the effective electron (gray line) in the first RF cycles shows some collisions with the top metallic or bottom dielectric surface that makes $N$ to increase, there is no exponential growth of the total number of electrons $N$, and consequently, no DC field is observed in the waveguide. In the following RF cycles, as the effective electron moves away from the center of the guide to positions of lower $E_{y,\text{RF}}$ (red line) due to the Miller force \cite{ar:Miller}, there are only low impact energy collisions, progressively decreasing $N$, until the discharge finally turns off. 
\begin{figure}[htbp]
	\centerline{\includegraphics[width=0.8\textwidth]{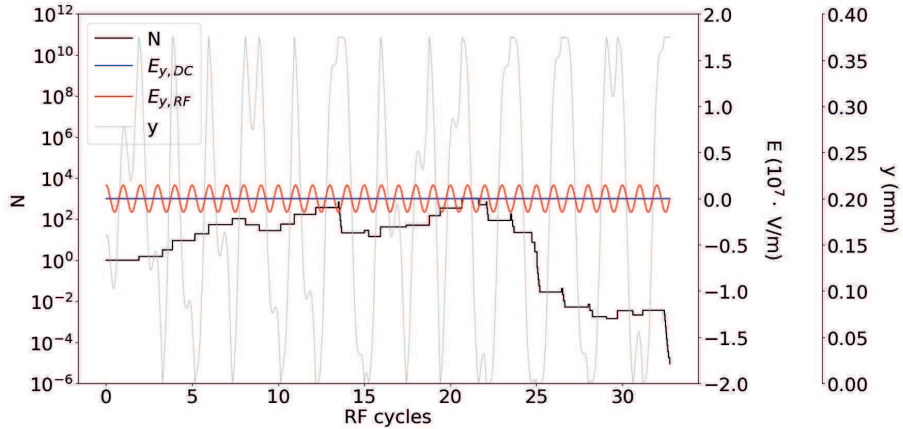}}
	\caption{Multipactor evolution in the partially dielectric-loaded rectangular waveguide at $f\times d=4.01$\,GHz$\cdot$mm and $V_\text{eff}= 550$\,V: total number of electrons $N$ (black line), $E_{y,DC}$ at the electron position (blue line), $E_{y,\text{RF}}$ (red line) and electron position $y$ (gray
line).}\label{f:rect550}
\end{figure}
A different result has been observed for this $V_\text{eff}$ and $f \times d$ point in the partially dielectric-loaded parallel-plate waveguide, in which 3 out of 360 simulations have shown a net growth of the population of electrons after 100 RF cycles, yielding an arithmetic mean of the final population of electrons using all simulations greater than~1 (see that this point is depicted in gray in Fig. \ref{fig:carta} for the parallel-plate waveguide, corresponding to a multipactor point). In Fig. \ref{f:gppp550} it is represented the multipactor evolution in one of such simulations, in which there is no longer a 1st order multipactor resonance regime at the sight of the $y$ coordinate of the effective electron, and consequently, there is a slower growth of the population of $N$ (in any case, it is not exponential), achieving a final high value of $10^{8}$ electrons at the end of the simulation. In the same way, there is no signiticant DC field appearing in the waveguide in this simulation (it is several orders of magnitude lower than the RF field), and added to the fact that the RF field in this waveguide does not depend on the $x$ coordinate, this regime of progressive increase of $N$ is not broken throughout the simulation. Thus, it can be concluded that a different evolution of the multipactor is observed in this point in the lower limit of the first order multipactor: an initial growth of the population of electrons in the rectangular waveguide in the first RF cycles, ending in the turning off of the discharge due to the Miller force effect, while in the parallel-plate waveguide a slow but progressive increase of the population of electrons allover the
simulation associated to the uniformity of the RF and DC fields in this waveguide.
coordinate.
\begin{figure}[htbp]
	\centerline{\includegraphics[width=0.8\textwidth]{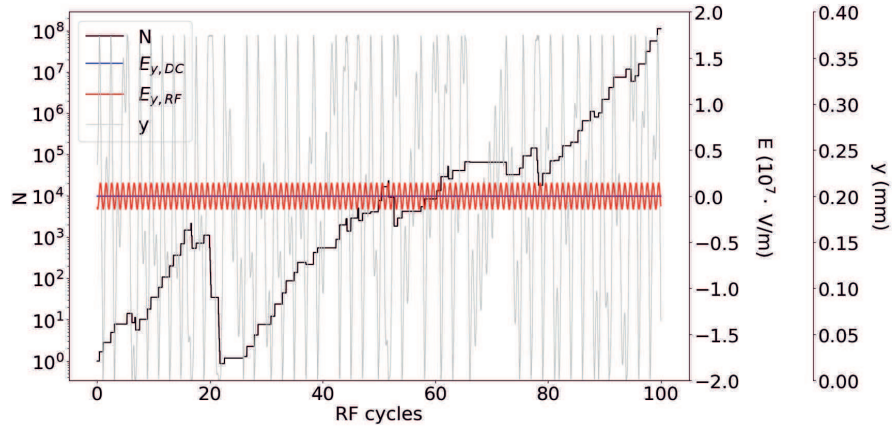}}
	\caption{Multipactor evolution in the partially dielectric-loaded parallel-plate waveguide at $f\times d=4.01$\,GHz$\cdot$mm and $V_\text{eff}= 550$\,V: total number of electrons $N$ (black line), $E_{y,DC}$ at the electron position (blue line), $E_{y,\text{RF}}$ (red line) and electron position $y$ (gray line).}\label{f:gppp550}
\end{figure}

\psection{Conclusion}
In order to highlight the similarities and differences between the multipactor results obtained in a partially dielectric-loaded rectangular waveguide and in its equivalent partially dielectric-loaded parallel-plate waveguide with the same height, a comparative study of the susceptibility chart in both cases has been performed. The conclusion of this study is that the inhomogeneity of the electric field inside the rectangular waveguide basically modifies the edges of the multipactor region in the susceptibility chart with respect to those predicted with the parallel-plate waveguide simulation code. In order to get a proper insight of this result, the evolution of the population and the trajectory of the effective electron and the DC field appearing in the waveguide in a point well inside the multipactor region obtained with both models is shown, revealing some quantitative differences in the multipactor evolution, but a similar qualitative behaviour is observed. On the other hand, when choosing a different point of the susceptibility chart in the lower limit of the first order multipactor region, a different evolution of the multipactor is observed in both cases, i.e., an initial growth of the population of electrons in the rectangular waveguide in the first RF cycles, ending in the turning off of the discharge due to the Miller force effect, while in the parallel-plate waveguide we can see a slow but progressive increase of the population of electrons allover the simulation, associated to the fact that the RF field in this waveguide does not depend on the $x$ coordinate.

\ack
This work was supported by the Agencia Estatal de Investigaci{\'o}n (AEI) and by the Uni{\'o}n Europea through the Fondo Europeo de Desarrollo Regional - FEDER - “Una manera de hacer Europa” (AEI/FEDER, UE), under the Research Project TEC2016-75934-C4-2-R.

\end{paper}
%--------------------------


\begin{thebibliography}{99}

\bibitem{ar:Esteban} Esteban, H., Morro, J.V., Boria, V. E., Bachiller, C., San Blas, A., and Gil, J., ``Multipaction modelling of low-cost H-plane filters using an electromagnetic field analysis tool," in {\it IEEE Antennas and Propagation Society International Symposium}, San Antonio, TX, USA, June 2004, 2155-2158.

\bibitem{ar:Quesada} Quesada, F., Boria, V.E., Gimeno, B., Ca\~nete, D., Pascual, J., \'Alvarez, A., Hueso, J., Schmitt, D., Raboso, D., Ernst, C., and Hidalgo, I., ``Investigation of multipactor phenomena in inductively coupled passive waveguide components for space applications," in {\it IEEE MTT-S Digest}, San Francisco, CA, USA, 2006, 246--249.

\bibitem{ar:Hatch} Hatch, A. J., and Williams, H. B., ``The secondary electron resonance mechanism of low-pressure high-frequency gas breakdown," {\it JJ. Appl. Phys.\/}, Vol.~25, No.~4, 417--423, April 1954.

\bibitem{ar:Torregrosa06} Torregrosa, G., Coves, A., Vicente, C. P., P\'erez, A. M., Gimeno, B., and Boria, V. E., ``Time evolution of an electron discharge in a parallel-plate dielectric-loaded waveguide," {\it IEEE Electron Device Lett.\/}, Vol.~27, No.~7, 629--631, 2006.

\bibitem{ar:Coves08} Coves, A., Torregrosa, G., Vicente, C. P., P\'erez, A. M., Gimeno, B. and Boria, V. E. ``Multipactor discharges in parallel plate dielectric-loaded waveguides including space-charge effects," {\it IEEE Trans. Electron Devices\/}, Vol.~55, No.~9, 2505--2511, 2008.

\bibitem{ar:Torregrosa10} Torregrosa, G., Coves, A., Gimeno, B., Montero, I.,  Vicente, C. P., and Boria, V. E. ``Multipactor Susceptibility Charts of a Parallel-Plate Dielectric-Loaded Waveguide," {\it IEEE Trans. Electron Devices\/}, Vol.~57, No.~5, 1160--1166, 2010.

\bibitem{ar:berenguer19} Berenguer, A., Coves, A., Mesa, F., Bronchalo, E., and Gimeno, B., ``Analysis of multipactor effect in a partially dielectric-loaded rectangular waveguide," {\it IEEE Trans. Plas. Sci.\/}, Vol.~47, No.~1, 259--265, 2019.

\bibitem{ar:berenguer16} Berenguer, A., Coves, A., Mesa, F., Bronchalo, E., Gimeno, B., and Boria, V. E., ``Calculation of the electrostatic field in a dielectric-loaded waveguide due to an arbitrary charge distribution on the dielectric layer," in {\it Proc. PIERS}, Shanghai, China, August 2016, 3251--3255.

\bibitem{ar:berenguer15} Berenguer, A., Coves, A., Bronchalo, E., Gimeno, B., and Boria, V. E. ``Analysis of 
multipactor effect in parallel-plate and rectangular waveguides," in  {\it Proc. PIERS}, Prague, Czech Republic, July 2015, 1564--1568.

\bibitem{Harrington} 
Harrington, R. F.,  {\it Time-Harmonic Electromagnetic Fields}, Wiley-IEEE Press, 2001.

\bibitem{ar:Scholtz} Scholtz, J.J., Dijkkamp, D., and Schmitz, R.W.A. ``Secondary electron emission properties," {\it Philips J. Res.\/}, Vol.~50, 375--389, 1996.

\bibitem{ar:Miller} Gaponov, A.V., and Miller, M.A., ``Potential wells for charged particles in high-frequency electromagnetic field," {\it Sov. Phys. JETP\/}, Vol.~7, 168, 1958.

\end{thebibliography}
\end{document}